# Ontology Bulding vs Data Harvesting and Cleaning for Smart-city Services

Pierfrancesco Bellini, Paolo Nesi, Nadia Rauch
*DISIT Lab, Dep. of Information Engineering, University of Florence, Italy*
*http://www.disit.dinfo.unifi.it, {pierfrancesco.bellini, paolo.nesi, nadia.rauch}@unifi.it*

*Abstract*— Presently, a very large number of public and private data sets are available around the local governments. In most cases, they are not semantically interoperable and a huge human effort is needed to create integrated ontologies and knowledge base for smart city. Smart City ontology is not yet standardized, and a lot of research work is needed to identify models that can easily support the data reconciliation, the management of the complexity and reasoning. In this paper, a system for data ingestion and reconciliation smart cities related aspects as road graph, services available on the roads, traffic sensors etc., is proposed. The system allows managing a big volume of data coming from a variety of sources considering both static and dynamic data. These data are mapped to smart-city ontology and stored into an RDF-Store where they are available for applications via SPARQL queries to provide new services to the users. The paper presents the process adopted to produce the ontology and the knowledge base and the mechanisms adopted for the verification, reconciliation and validation. Some examples about the possible usage of the coherent knowledge base produced are also offered and are accessible from the RDF-Store.

*Keywords— Smart city, knowledge base construction, reconciliation, validation and verification of knowledge base, smart city ontology, linked open graph.*

## I. Introduction

Despite to the large work performed by Public Administrations, PAs, on producing open data they are not typically semantically interoperable and neither with the many private data. Open data coming from PA contains typically statistic information about the city (such as data on the population, accidents, flooding, votes, administrations, etc.), location of point of interests on the territory (including, museums, tourism attractions, restaurants, shops, hotels, etc.), major GOV services, ambient data, weather status and forecast, changes in traffic rules for maintenance interventions, etc. Moreover, a relevant role is covered in city by private data coming from mobility and transport such as those created by Intelligent Transportation Systems, ITS, for bus management, and solutions for managing and controlling parking areas, car and bike sharing, car flow, delivering organizations, accesses on Restricted Traffic Zone, RTZ, etc. They can include real time data such as the traffic flow measure, position of vehicles (buses, car/bike sharing, taxi, garbage collectors, delivering services, etc.), railway and train status, park areas status, and Bluetooth tracking systems for monitoring movements of cellular phones, ambient sensors, and TV cameras streams for security. Both PAs and mobility operators have large difficulties in elaborating and aggregating these data to provide new services, even if they could have a strong relevance in improving the citizens' quality of life. Therefore, our cities are not so smart as they could be by exploiting a semantically interoperable knowledge base founded on these data. This condition is also present in highly active cities on open data publication such as Firenze, that is considered one of the top cities on Open Data.

Therefore, the variability, complexity, variety, and size of these data, make the data process of ingestion and exploitation a big data problem as addressed in [2], [3]. The variety and variability of data can be due to the presence of different formats, and to scarce (or non-existing) interoperability among semantics of the single fields and of the several data sets. In order to reduce the ingestion and integration cost, by optimizing services and exploiting integrated information, a better interoperability and integration among systems is required [1], [2]. This problem can be partially solved by using specific reconciliation processes to make these data interoperable with other ingested and harvested data. The velocity of data is related to the frequency of data update, and it allows to distinguish static data from dynamic data: the first one are rarely updated, like once per month/year, as opposed to the second one that are updated once a day up to every minute or more. When these data models are analyzed and then processed to become semantically interoperable, they can be used to create a common knowledge base that can be feed by corresponding data instances (with static, quasi-static and real time data). This process may lead to create a large interoperable knowledge base that can be used to make queries for producing suggestions as well as, predictions, deductions, in the navigation or in the service access and usage.

This scenario enables the creation of new services exploiting the accumulated knowledge for: delivering service predictions and tuning, deducing and predicting critical conditions, towards different actors: public administrations, mobility operators, commercials and point of interests and citizens. In this paper, the above mentioned complex process of knowledge base construction is described from: ontology creation to the data ingestion and knowledge base production and validation. The mentioned process also include processes of data analysis for ontology modeling, data mining, formal verification of inconsistencies and incompleteness to perform data reconciliation and integration. Among the several processes, the most critical aspects are related to the ontology construction that can enable deduction and reasoning, and on the verification and validation of the obtained model and knowledge base.

The paper is organized as follows. In Section II, the overview of the proposed ontology is present together with the main problems underlined its construction, and the main macro classes. Section III describes the details associated to each macroclass of the proposed smart city ontology and the integration with other vocabulary. In Section IV, the general architecture adopted for processing Open Data and the motivations that constrained its definition are reported. Section V presents the verification and validation process adopted to produce and verify the knowledge base. In the same section, two services are presented that allow navigating in the knowledge and can be used by non-data engineers to inspect and navigate into the knowledge base. Conclusions are drawn in Section VI.

## II. Ontology main elements

In order to create an ontology for Smart City services, a large number of data sets have been analyzed to see in detail each single data elements of each single data set with the aim of modeling and establishing the needed relationships among element, thus making a general data set semantically interoperable (e.g., associating the street names with toponimous coding, resolving ambiguities). The work performed started from the data sets available in the Florence and Tuscany area. In total the whole data sets are more than 800 data sets. At regional level, Tuscany Region also provided a set of open data into the MIIC (Mobility Integration Information Center of the Tuscany Region), and provide also integrated and detailed geographic information reporting each single street in Tuscany (about 137,745), and the location of a large part of civic numbers, for a total of 1,432,223 (a wider integration could be performed integrating also Google maps and Yellow/white pages). From the MIIC it is possible to recover information regarding streets, car parks, traffic flow, bus timeline, etc. While from Florence municipality real time data about the RTZ, tram lines on the maps, bus stops, bus tickets, accidents, ordinances and resolutions, numbers of arrivals in the city, number of vehicles per year, etc. From the other open data points of interest can be recovered as position and information related to: museums, monuments, theaters, libraries, banks, express couriers, police, firefighters, restaurants, pubs, bars, pharmacies, airports, schools, universities, sports facilities, hospitals, emergency rooms, government offices, hotels and many other categories, including weather forecast by Lamma consortium. In addition to these data sets, those coming from the mobility and transport operators have been collected as well.

The analysis of the above mentioned data sets allowed us to create an integrated ontological model presenting 6 main areas of macroclasses as depicted in Figure 1.

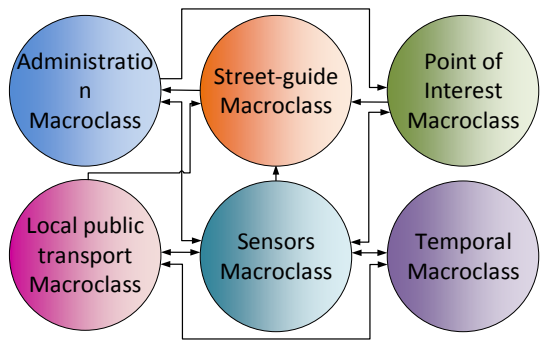

**Figure 1 - Ontology Macro-Classes and their connections**

*Administration*: includes classes related to the structuring of the general public administrations, namely PA, and its specifications, Municipality, Province and Region; also includes the class Resolution, which represents the ordinance resolutions issued by each administration that may change the traffic stream.

*Street-guide*: formed by entities as Road, Node, RoadElement, AdminidtrativeRoad, Milestone, StreetNumber, RoadLink, Junction, Entry, and EntryRule Maneuver, it is used to represent the entire road system of Tuscany, including the permitted maneuvers and the rules of access to the limited traffic zones. The street model is very complex since it may model from single streets to areas, different kinds of crosses and superhighways, etc. In this case, OTN vocabulary has been exploited to model traffic [4] that is more or less a direct encoding of GDF (Geographic Data Files) in OWL.

*Point of Interest*: includes all services, activities, which may be useful to the citizen and who may have the need to search for and to arrive at. The classification of individual services and activities is based on main and secondary categories planned at regional level. In addition, this macro segment of the ontology may take advantage of reusing Good Relation model of the commercial offers[1].

*Local public transport*: includes the data related to major TPL (Transport Public Local) companies scheduled times, the rail graph, and data relating to real time passage at bus stops. Therefore this macroclass is formed by classes TPLLine, Ride, Route, AVMRecord, RouteSection, BusStopForeast, Lot, BusStop, RouteLink, TPLJunction.

*Sensors*: macroclass concerns data from sensors: ambient, weather, traffic flow, pollution, etc. Currently, data collected by various sensors installed along some streets of Florence and surrounding areas, and those relating to free places in the main car parks of the region, have been integrated in the ontology.

*Temporal*: macroclass that puts concepts related to time (time intervals and instants) into the ontology, so that associate a timeline to the events recorded and is possible to make forecasts. It may take advantage from time ontologies such as OWL-Time [5].

The ontology reuses the following vocabularies: *dcterms*: set of properties and classes maintained by the Dublin Core Metadata Initiative; *foaf*: dedicated to the description of the relations between people or groups; *vCard*: for a description of people and organizations; *wgs84_pos*: vocabulary representing latitude and longitude, with the WGS84 Datum, of geo-objects.

III. SMART-CITY ONTOLOGY DETAILS

A. *Administration Macroclasss*

The Administration Macroclass is structured in order to represent the Italian public administration hierarchy: each region is divided into several provinces, within which the territory is again divided into municipalities. Moreover each PA, during its mandate, can produce resolutions and publish statistics. To represent this situations the SmartCity Ontology has, as main class of Administration Macroclass, the class *Pa*, which has been defined as a subclass of *foaf:Organization*, link that helps to assign a clear meaning to this class. The three subclasses of *Pa*, i.e. *Region*, *Province* and *Municipality* are automatically defined according to the restriction on some ObjectProperties: for example, the class *Region* is defined as a restriction of the class *PA* on ObjectProperty *hasProvince*, so that only the PA that possess provinces, can be classified as *Regions*. The class *PA* is connected to the *Resolution* class through the ObjectProperty *hasApproved*, that has its inverse property, *approvedBy*. Statistical data related to both various municipalities in the region and to each street, are represented by a unique class *StatisticalData*, shared by the macroclasses Administration and Street Guide: as we will see also in the next subsection, the class *StatisticalData* is connected to both classes *Pa* and *Road* through the ObjectProperty *hasStatistic*.

B. *Street-guide Macroclass*

At regional level, the entire roads system in Tuscany, from an administrative point of view, is seen as a set of administrative extensions or administrative roads, while from the citizen' point of view, it is composed by a set of roads. Each administrative road represents the administrative division of the roads, based on which PA have to manage them. Both administrative roads and roads are formed by a variable number of road elements, each of which starts and ends in a unique node. Each road element, in turn, is formed by a set of sections separated by an initial junction and a final junction, which allow to delineate the exact broken line that represents each road element. Placed on the various roads there are street numbers, each of which always corresponds to at least one entry; in some cases there are two entrances which corresponds to a single street number, i.e. the outer gate and the front door. With regard to road circulation, access rules and maneuvers are defined: the first one defines access restrictions to each road element, the second one, instead, are mandatory turning maneuvers, priority or forbidden, which are described by indicating the order of road elements involving.

---

[1] http://www.heppnetz.de/projects/goodrelations/

Another element present into the Tuscany road system is the milestone, which represents the kilometer stones that are placed along the administrative roads, that is, the elements that identify the precise value of the mileage at that point.

The situation described above has been modeled into the Smart City Ontology, choosing as the main class of Street Guide macroclass, the *RoadElement* class, which is defined as a subclass of the corresponding element in the OTN Ontology (see Figure 2), that is *Road_Element*. Each road element is delimited by a start node and an end node, detectable by the ObjectProperties *starts* and *ends*, which connect elements of the class in question to the class *Node*, subclass of the same name class *OTN:Node*, belonging to ontology OTN.

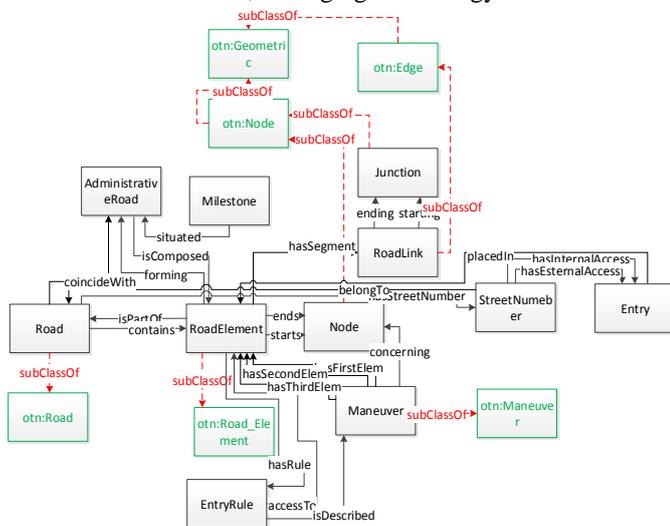

**Figure 2 - The Street-guide Macro class**

The class *Node* has been defined with a restriction on DataProperty *geo:lat* and *geo:long*, two properties inherited from the definition of the class *Node* as subclass of *geo:SpatialThing* belonging to ontology Geo wgs84 [7]: in fact, each node can be associated with only one pair of coordinates in space, and cannot exist a node without these values. The class *Road* is defined as a subclass of the corresponding class in the OTN Ontology, i.e., the homonymous class *Road*, with a cardinality restriction on the ObjectProperty *contains*, since a road that does not contain at least one road element, cannot exist. Also the class *AdministrativeRoad* is connected to class *RoadElement* through two inverse ObjectProperties *isComposed* and *forming*, while it is connected with only one ObjectProperty, *coincideWith*, to the class *Road*. In order to better clarify the relationship that exists between classes *Road*, *AdministrativeRoad* and *RoadElement*: a *Road*'s instance can be connected to multiple instances of class *AdministrativeRoad* (e.g., if a road crosses the border between two provinces), but the opposite is also true (e.g., when a road crosses a provincial town center and it assumes different names), i.e., there is a N:M relationship between these two classes. On each road element, it is possible to define access restrictions, identified by the class *EntryRule*, which is connected to the class *RoadElement* through 2 inverse ObjectProperties, i.e., *hasRule* and *accessTo*. The class *Manoeuvre* and the class *EntryRule* are connected by the ObjectProperty *isDescribed*; moreover verified that only in rare cases maneuvers involving three different road elements, to represent the relationship between classes *Maneuvre* and *RoadElement*, three ObjectProperties were defined: *hasFirstElem*, *hasSecondElem* and *hasThirdElem*, in addition to the ObjectProperty that binds a maneuver to the junction that is interested, that is, *concerning* (because a maneuver takes place always in proximity of a node). Each instance of *Milestone* class must be associated with a single instance of *AdministrativeRoad*, and it is therefore defined a cardinality restriction equal to 1, associated to the ObjectProperty *placedIn*; also the class *Milestone* is defined as subclass of *geo:SpatialThing*, but this time the presence of coordinates is not mandatory. Thanks to the owned data, the classes *StreetNumber* and *Entry* were then defined: the connection of *StreetNumber* class to the class *RoadElement* and to the class *Road*, is possible respectively through the ObjectProperties *standsIn* and *belongTo*; the relationship between the classes *Entry* and *StreetNumber*, is also defined by two ObjectProperties, *hasInternalAccess* and *hasExternalAccess*. The class *Entry* is also defined as a subclass of *geo:SpatialThing*, and it is possible to associate a maximum of one pair of coordinates *geo:lat* and *geo:long* to each instance. The Street-guide macroclass is connected to the Administration macroclass through two different ObjectProperties, i.e. *OwnerAuthority* and *managingAuthority*, which as the name suggests, clearly represent respectively the public administration which owns an *AdministrativeRoad*, or public administration that manages a *RoadElement*. Thanks to the processing of *KMZ* files, is possible to retrieve the set of coordinates that define the broken line of each *RoadElement*, and each of these points will be added to the ontology as an instance of class *Junction* (defined as a subclass of *geo:SpatialThing*, with compulsory single pair of coordinates). Each small segment between two instances of *Junction* class is instead an instance of the class *RoadLink*, which is defined by a restriction on the ObjectProperties *ending* and *starting*, which connect the two mentioned classes. RoadLink and Juctions are additional 20 million of triples.

### C. Point of Interest Macroclass

This macroclass allows to represent services to the citizens, points of interest, businesses activities, tourist attractions, and anything else can be located thanks to a pair of coordinates on a map. Each different type of elements has been defined starting from the categories defined by the Tuscany Region: Accommodation, GovernmentOffice, TourismService, TransferService, CulturalActivity, FinancialService, Shopping, Healthcare, Education, Entertainment, Emergency and WineAndFood.

It is easy to understand that the main class of the Point of Interest Macroclass is a generic class *Service* for which the subclasses above listened have been identified thanks to the value assigned to the ObjectProperty *serviceCategory*.

The class *Accommodation* for example, was defined as a restriction of the class *Service* on the ObjectProperty *serviceCategory*, which must take one of the following values: *villaggio_vacanze*, *albergo_hotel*, *casa_per_vacanze*, *casa_di_riposo*, *casa_per_ferie*, *bed_and_breakfast*, *hostel*, *residenza_turistica_alberghiera*, *residence_di_villeggiatura*, *farmhouse*, *centri_accoglienza_ e_case_alloggio*, *camping*, *residenze_epoca*, *rifugio_alpino*.

We have also defined the DataProperty ATECOcode, i.e. the ISTAT code for classification of economic activities, which could be used in future as a filter to define the various services subclasses, in place of the categories proposed by the Tuscany Region, in order to make more precise research of the various types of services. Thanks to the class *Service* the macroclasses *Point of Interest* and *Street guides* can be connected by exploiting the ObjectProperty *hasAccess*, with which a service can be connected to only one external access, corresponding to the road and the street number of the service location. If this association is not possible (because of lack of information, missing street number, etc..), the connection between the same two macroclasses listed above, is realized through the ObjectProperty *isIn*, that connects an instance of the class *Service* to an instance of the class *Road*. In order to use at least one of these two ObjectProperty to connect the macroclasses *Point of Interest* and *Street Guides*, an intense reconciliation phase is necessary, as described in *section IV*.

### D. Transport Public Local Macroclass

The TPL (Transport Public Local) macroclass (see Figure 3) includes information relating to public transport by road and rail. The public transport by road is organized in public transport lots, each of which is in turn composed of a number of bus/tram lines. Each line includes at least two ride (the first in ascendant direction, and the second one in descendant direction), identified through a code provided by the TPL company and each ride is scheduled to drive along a specific path, called route. A route can be seen as a series of road segments delimited by subsequent bus stops, but wishing then to represent to a cartographic point of view the path of a bus, we need to represent the broken line that composes each stretch of road crossed by the means of transport itself, and to do so, the previously used modeling on road elements, has been reused: we can see each path as a set of small segments, each of which delimited by two junctions.

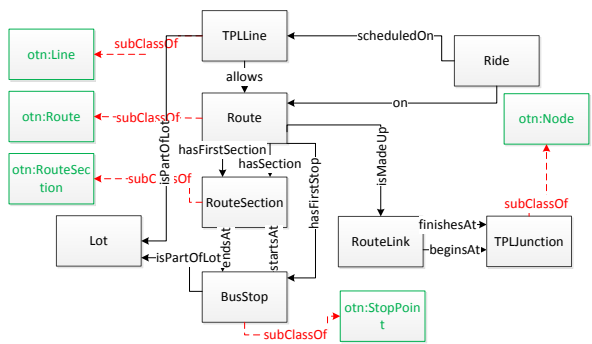

**Figure 3 - Public Transport Macroclass (a portion)**

The part relating to rail transport is significantly easier: each railway line, i.e. an infrastructure designed to run trains between two places of service, is composed by a number of railway elements, which can also form a railway direction (a railway line having particular characteristics of importance for volume of traffic and transport relations linking centers or main nodes of the rail network) and a railway section (section of the line in which you can find only one train at time, and that is usually preceded by a "protective" or "block" signal). In addition, each rail element begins and ends at a railway junction, in correspondence of which there may be train stations or cargo terminals.

Based on the previously written, we have defined the class *TPLLine* (that it is also subclass of *OTN:Line*), which is connected to the corresponding instance of the class *Lot*, thanks to the ObjectProperty *isPartOfLot*. Every instance of the class *TPLLine* is connected to the class *Ride* through the ObjectProperty *scheduledOn*, which is also defined as a limitation of cardinality exactly equal to 1, because each stroke may be associated to a single line. To model each path and its sequence of crossed bus stops, the classes *Route* and *BusStop* have been defined: we decided to define two ObjectProperties linking the classes *Route* and *RouteSection*, i.e. *hasFirstSection* and *hasSection*, since, from a cartographic point of view, wanting to represent the path that a certain bus follows; knowing the first segment and the stop of departure, it is possible to obtain all the other segments that make up the complete path and, starting from the second bus stop, that is identified as the different stop from the first stop, but that it is also contained in the first segment, we are able to reconstruct the exact sequence of the bus stops, and then the segments, which constitute the entire path. For this purpose also the ObjectProperty *hasFirstStop* has been defined, which connects the classes *Route* and *BusStop* and the ObjectProperty *endsAt* and *startsAt*, which connect instead each instance of *RouteSection* to two instances of the class *BusStop* (subclass of *OTN:StopPoint*). Each stop is also connected to the class *Lot*, through the ObjectProperty *isPartOfLot*, with a 1:N relation, because there are stops shared by urban and suburban lines so they belong to two different lots. Possessing also the coordinates of each stop, the class *BusStop* was defined as a subclass of *geo:SpatialThing*, and was also termed a cardinality equal to 1 for the two DataProperty *geo:lat* and *geo:long*. To represent the broken line that composes each route the classes *RouteLink* and *TPLJunction*, and the ObjectProperties *beginsAt* and *finishesAt*, were defined. The class *Route* is also connected to the class *RouteLink* through *isMadeUp* ObjectProperty.

The Railway Graph is mainly formed by the class *RailwayElement*, that can be connected to the classes *RailwayDirection* and *RailwaySection*, thanks to two inverse ObjectProperties *isComposedBy* and *composing*, and to the class *RailwayLine*, trough the two inverse ObjectProperties *isPartOfLine* and *hasElement*. Each instance of the class *RailwayElement* is connected to two instances of the class *RailwayJunction* (defined as a subclass of the OTN:Node), by the ObjectProperties *startAt* and *endAt*,. The classes *TrainStation* and *GoodsYard* correspond only to one instance of the *RailwayJunction* class, both through the ObjectProperty *correspondTo*.

### E. Sensors Macroclass

Sensors Macroclass has not yet been completed, but for now it consists of four parts respectively relating to car parks sensors, weather sensors, traffic sensors installed along roads/rails and to AVM systems installed on buses.

The first part is focused on the real-time data related to parking: for each sensors installed into different car parks, we receive a situation record every 5minutes, in which there are information about the number of free and occupied parking spaces, for the main car parks in Tuscany Region.

The second part of the received real-time data, concerns the weather forecast, thanks to LAMMA. This consortium will update each municipality report once or twice a day and every report contains forecast of five days divided into range, which have a greater precision (and a higher number) for the nearest days until you get to a single daily forecast for the 4th and 5th day.

The third part of the real-time data concerns the sensors placed along the roads of the region, which allow making different detection related to traffic situation. Unfortunately, the location of these sensors is not very precise, it is not possible to place them in a unique point thanks to coordinate, but only to place them within a toponym, which for long-distance roads such as FI- PI-LI road, it represents a range of many miles. Each sensor, is part of a group and produces observations which can belong to four types, i.e. they can be related to the average velocity, car flow passing in front of the sensor, traffic concentration, or to the traffic density.

The last part of Sensors macroclass concerns the AVM (Authomatic Vehicle Monitoring) systems installed on most of ATAF buses, which, at intervals of few minutes, send a report to the management center, and they contains the following information: the last stop done, GPS coordinates of the vehicle position, the identifiers of vehicle and line, a list of upcoming stops with the planned passage time.

To model the car parks situation we have defined the class *CarParkSensor* which is linked to instances of the class *SituationRecord*, that represent, as previously stated, the state of a certain parking at a certain instant; this link is performed via the reverse ObjectProperties, *relatedTo* and *hasRecord*. This first part of the Sensors Macroclass is also connected to the Point of Interest Macroclass through two inverse ObjectProperties, *observe* and *isObservedBy*, which connect the classes *CarParkSensor* and *TransferService*.

The weather situation, instead, is represented by the class *WeatherReport* connected to the class *WeatherPrediction* via the ObjectProperty *isComposedOf*. Moreover, the class *Municipality* is connected to each report by two reverse ObjectProperties: *refersTo* and *has*, to realize the connection between the macroclasses Sensors and Administration.

With regard to traffic sensors, each group of sensors is represented by the class *SensorSiteTable* and each instance of the class *SensorSite* is connects to its group through the ObjectProperty *forms* and thanks to the ObjectProperty *installedOn* each instance of the class *SensorSite* can be connected only to the class *Road* (see Figure 4), to create a connection between Sensors and Street-guide macroclasses. Each sensor produces observations represented by instance of class *Observation* and, as mentioned earlier, there are four possible subclasses: *TrafficConcentration, TrafficHeadway, TrafficSpeed*, and *TrafficFlow* subclass. The classes *Observation* and *Sensor* are connected via a pair of reverse ObjectProeprties, *hasProduced* and *measuredBy*.

Finally, the last part of Sensors Macroclass is mainly represented by two classes, *AVMRecord* and *BusStopForecast*, and thanks to the ObjectProperty *lastStop*, this first class can be connected to the *BusStop* class. The list of scheduled stops is instead represented as instances of the class *BusStopForecast*, a class that is linked to the class *BusStop* through *atThe* ObjectProperty so as to be able to recover the list of possible lines provided on a certain stop (the class *AVMRecord* is in fact also connected to the class *Line* via the ObjectProperty *concern*).

### F. Temporal Macroclass

Finally, the last macroclass, called Temporal Macroclass, is now only "sketchy" within the ontology, and it is based on the Time ontology [5] as it has been used into OSIM ontology [8]. It requires the integration of the concept of time as it will be of paramount importance to be able to calculate differences between time instants, and the Time ontology comes to help us in this task. We define fictitious URI *#instantForecast*, *#instantAVM*, *#instantParking*, *#instantWreport*, *#instantObserv* to following associate them to the identifier URI of a resource referred to the time parameter, i.e. respectively *BusStopForecast*, *AVMRecord*, *SituationRecord*, *WheatherReport* and finally *Observation*. It is necessary to create a fictitious URI that links a time instant to each resource, to not create ambiguity, because identical time instants associated with different resources may be present (although the format in which a time instant is expressed has a fine scale). Time Ontology is used to define precise moments as temporal information, and to use them as extreme for intervals and durations definition, a feature very useful to increase expressiveness.

Pairs of ObjectProperties have also been defined for each class that needs to be connected to the class *Instant*: between classes *Instant* and *SituationRecord* were defined the inverse ObjectProperties *instantParking* and *observationTime*, between classes *WeatherReport* and *Instant*, the ObjectProperties *instantWReport* and *updateTime* have been defined; between classes *Observation* and *Time* there are the reverse ObjectProperties *measuredTime* and *instantObserv*, between *BusStopForecast* and *Time* we can find *hasExpectedTime* and *instantForecast* ObjectProperties, and finally, between *AVMRecord* and *Time*, there are the reverse ObjectProperties *hasLastStopTime* and *instantAVM*.

### IV. DATA ENGINEERING ARCHITECTURE

In this section, the description of the data engineering architecture is proposed (Figure 4). The process can be divided into four phases: Data **Ingestions**, knowledge **Mapping**, and interoperable knowledge **Validation** and **Access**/exploitation from services. The set of ingestion processes is managed by a **Process Scheduler** that allocates processes on a parallel and distributed architecture composed by several servers. To allow the regular update of ingested data the scheduler regularly retrieves data and check for updates. The ingested data are transcoded and then mapped in the DISIT Ontology for Smart City. After that, they are made available to applications on an RDF Store (OWLIM-SE) using a SPARQL Endpoint. Applications can use the geo-referenced data to provide advanced services to the city citizens, such as the present solution for knowledge base browsing via Linked Open Data (http://log.disit.org) and the Service Map (Http://servicemap.disit.org), described in the following.

### A. Data Ingestion

For the data ingestion, the problems are related to the management of the several formats and of the several data sets that may find allocation on different segments and areas of the Smart City Ontology. The solution allows ingesting and harvesting a wide range of public and private data, coming as static, semi-static and real time data as mentioned in the previous sections. For the case of Florence area, we are addressing about 150 different sources of the 564 available. **Static and semi-static data** include points of interests, geo-referenced services, maps, accidents statistic, etc. This information is typically accessible as public files in several formats, such as: SHP, KML, CVS, ZIP, XML, etc.

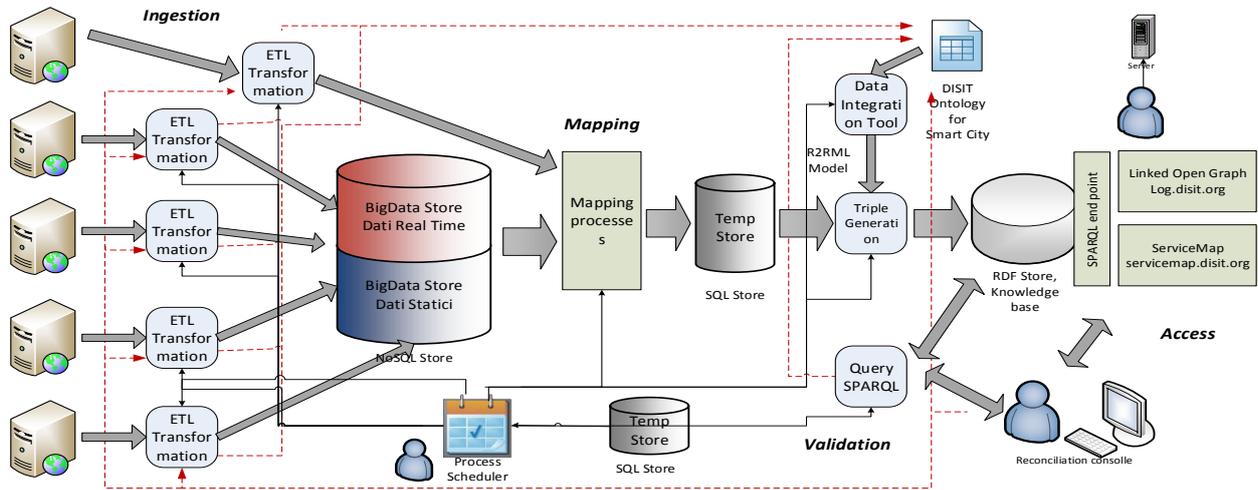

Figure 4 - Architecture Overview

Each Open Data ingestion process retrieves information and produce records in a noSQL Hbase for bigdata [9], logging all the information acquired to trace back and versioning the data ingestion. Data are then completed, other columns are updated dynamically with other process steps, and finally data obtained are placed on an HBase table.

**Real time data** includes data coming from sensors (e.g., parking, weather conditions, pollution measures, busses, etc.) that are typically acquired from Web Services as well as more static data as road graph description, etc. For example ingestion of data relating to traffic sensors consists of a ETL transformation. In most cases, the real-time data are directly pushed in the mapping process to feed the temporary SQL store. They are typically streamed into the traditional SQL store and then converted into triples in the RDF final store.

In almost all cases, each single data set is ingested by means of a different ETL process defined by using Pentaho Kettle formalism [10] because, among the several existing solutions, this formalism is quite diffused and easier to understand, and it was already used by Information Systems Directorate of Florence. When the Kettle language presented limitation, external processes in Java have been adopted.

*B. Data Mapping*

The Mapping Phase deals with the transport of information, previously saved into HBase database, into an RDF datastore, in our case managed by Owlim-SE [11]. The first part of this procedure retrieves information from HBase to put them on a temporary MySQL database (required to use the Data Integration tool chosen), while in the second part data are translated into triples. Transformation is needed to map the traditional structured into RDF triples, based on information contained in a well-defined ontology (DISIT Ontology for Smart City) and all ontologies reused (dcterms, foaf, vCard, etc.). This process may be performed by ad-hoc programs that have to take into account the mapping from linear model to RDF structures. This two steps process allowed us to test and validate several different solutions for mapping traditional information into RDF triples and ontology. The ontological model has been several times updated and thus the full RDF storage has been regenerated from scratch reloading the definition (all the other vocabularies, selecting the testing several different solutions) and the instance triples according to the new model under test. Once the model has been generated, triples can be automatically inserted.

The first essential step is to specify semantic types of the data set, i.e., it is necessary to establish the relationship between the columns of the SQL tables and properties of ontology classes. The second step consists in defining the Object Properties among the classes, or the relationships between the classes of the ontology. When dataset has 2 columns that have the same semantic type but which correspond to different entities, thus multiple instances of the same class have to be defined, associate each column to the correct one.

The process responsible to perform the mapping transformation, passing from Hbase to SQL database has been produced as a corresponding ETL Kettle associated with each specific ingestion procedure for each data set. The second phase, of performing the mapping from SQL to RDF, has been realized by using a mapping model:Karma Data Integration tool [12], which generates a R2RML model, representing the mapping for transport from MySQL to RDF and then it is uploaded in a OWLIM-SE RDF Store instance [11]. Karma initialization phase involves loading the primary reference ontology and connecting dataset containing the data to be mapped. This process allowed the production of the knowledge base that may present a large set of problems due to inconsistencies and incompleteness that may be due to lack of relationships among different data sets, etc. These problems may lead to the impossibility of making deductions and reasoning on the knowledge base, and thus on reducing the effectiveness of the model constructed. These problems have to be solved by using a reconciliation phase via specific tools and the support of human supervisors.

*C. Data Reconciliation*

To connect services to the Street Guide in the repository a reconciliation phase in more steps, has been required, because the notation used by the Tuscany region in some Open Data within the Street Guide, does not always coincide with those used inside Open Data relating to different points of interest. In substance, different public administration are publishing Open Data that are not semantically interoperable.

Furthermore, there are different types of inconsistencies within the various integrated dataset, such as:
- typos;
- missing street number, or replacement with "0" or "SNC";
- Municipalities with no official name (e.g. Vicchio/Vicchio del Mugello);
- street names with strange characters ( -, /, ° ? , Ang., ,);
- street numbers with strange characters ( -, /, ° ?, Ang. ,(, );
- road name with words in a different order from the usual ( e.g. Via Petrarca Francesco, exchange of name and surname);
- number wrongly written (e.g. 34/AB, 403D, 36INT.1);
- red street numbers (in some cities, street numbers may have a color. So that a street may have 4/Black and 4/Red, red is the numbering system for shops);Roman numerals in the road name (e.g., via Papa Giovanni XXIII).

Thanks to the created ontology, is possible to perform services reconciliation at street number level, i.e. connecting an instance of class *Service* to an external access that uniquely identifies a street number on a road, or only at street-level, with less precision (lack that can be compensated thanks to geolocation of the service).

The methodology used in this reconciliation phase consists of first try to connect each service at street number-level, and then, perform the reconciliation at street-level.

The first reconciliation step performed consists of an exact search of the street name associated to each service integrated. For example, to reconcile the service located at "VIA DELLA VIGNA NUOVA 40/R-42/R, FIRENZE", a SPARQL query is necessary, to search for all elements of *Road* class connected to the municipality of *"FIRENZE"* (via the ObjectProperty *inMunicipalityOf*), which have a name that exactly corresponds to "VIA DELLA VIGNA NUOVA" (checking both fields: official name, alternative name). The query results has to be filtered again, imposing that an instance of *StreetNumber* class exists and it corresponds to civic number "40" or "42", with the R class code Red.

From this first reconciliation step, the services for which was identified a single instance of the class *Entry* has been selected, and the related reconciliation triples at street number-level, have been created.

A very frequent problem for exact search, is the existence of multiple ways to express toponym qualifiers, that is dug (e.g. Piazza and P.zza) or parts of the proper name of the street (such as Santa, or S. or S or S.ta): thanks to support tables, inside which the possible change of notation for each individual case identified are inserted, a second reconciliation step was performed, based on exact search of the street name, which has allowed to increase the number of reconciled services at street number-level.

The third reconciliation step is based on the research of the last word inside the field *v:Street-Address* of each instance of the *Service* class, because, statistically, for a high percentage of street names, this word is the key to uniquely identify a match.

These first three reconciliation steps have been also carried out without taking into account the street number, and so in order to obtain a reconciliation at street-level of each individual service.

The fourth reconciliation step is realized thanks to Google Geocoding API[2], through which different services, not yet connected to the *Street Guide* macroclass at street number-level, were searched again.

The next reconciliation step used automated methods to remove strange characters, inside the street number field, or the address field, but unfortunately at this point it is becoming increasingly difficult to obtain unique results in the search for correspondences between instances of the class *Entry* and instances of the class *Service*.

The last reconciliation step implemented, trying to reconcile all those services in which the name of the town is incorrectly used or it is expressed in a not official notation; even in this case it is difficult to get great results from every single reconciliation step.

At present, all services that present typos, street number equal to "0" or to string "SNC", still need to be managed; moreover services with strange characters in the street name, are partially managed.

As a summary, the whole knowledge base created at the first day has been of more than 81 Million triples, when it grows of 4 Million per month. A part of them can be discharged when statistical values are estimated and punctual value discharged. For the validation, a total amount of services/points of interest inserted into the repository has been of 30182 instances. Among these, 13185 have been reconciled at street number-level, while the number of elements reconciled at street-level has been 21207. There are also 149 services associated to a coordinate pair, for which reconciliation did not return any results, yet for the lack of references into the knowledge base (some streets and civic numbers are still missing or incomplete).

## V. VERIFICATION AND VALIDATION

The validation process is performed by defining a set of SPARQL queries that verify the knowledge base conditions with the aim of detecting inconsistencies and incompleteness, and verifying the correct status of the model. These queries have to be periodically executed in order to perform a regression testing every time a new update of data process ingestion is performed. So that, processes for ingestion and mapping have to be connected to validation processes that have to be re-executed. The validation process may lead to identify changes in the ingested data sets that may implies to apply changes into the ontological model or in the above mentioned processes. So that, an iterative workflow process is defined. In order to validate the ingestion performed a set of SPARQL queries were used. During validation there were cases like the Weather forecast where no connection among the data were present due to different encoding of the name of the municipality, for this reason to support the reconciliation process a table containing the ISTAT code of each municipality was created, and each time new weather data are available, they are automatically completed with the correct ISTAT code, thus supporting the search for the instance of the PA class to which connect the weather forecasts.

Following the reconciliation activity related to service, a series of SPARQL queries to verify the correctness of triples created, was performed. In some cases a manual intervention was necessary to select the corresponding instance of *Road* class and then create a triple owl:sameAs to the selected toponym.

---

[2] https://developers.google.com/maps/documentation/geocoding/

The system has been used to ingest the data coming from the Municipality of Florence, the Tuscany Region and MIIC. Considering only files related to the daily weather forecast of all the available municipalities, we have 286 files updated twice a day, each of which, containing also 16 lines of weather prediction for the week, we obtain an increase of approximately 270,000 HBase lines per month that, in terms of triples, corresponds to a monthly increase of about 2.5 million triples.

Moreover, in order to explore the data being ingested and their relationships a tool for data visualization and exploration was used, that allows exploring the semantic graph of the relations among the entities, this Linked Open Graph is available for applications developers to explore and understand better the data available in the ontology.

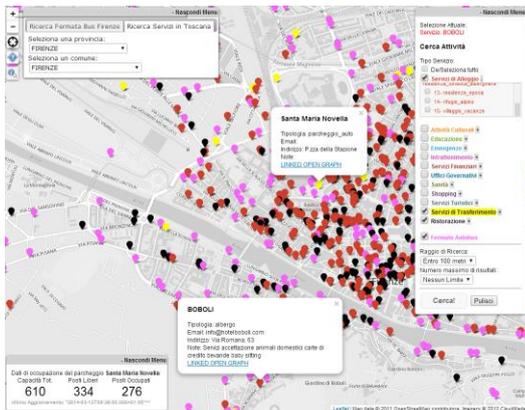

**Figure 5 - Service Map (http://servicemap.disit.org)**

A second tool called ServiceMap to perform geographic queries (for example to get points of interests close to a bus station, to get the street number close to a give point on the map, etc.) has been realized (see Figure 5).

The service map, for example, allows to (i) get bus stops and from them to access at the status line of the bus, providing the time to wait for the next bus, (ii) finding parking and getting in real time the number of empty places, etc. From each pin, it is possible to pass from the entity identified to its model in terms of relationships on the LOG graph.

## VI. CONCLUSIONS

In this paper, a system for the ingestion of public and private data for smart city with related aspects as road graph, services available on the roads, traffic sensors etc., has been proposed. The system includes both open data from public administration and private data coming from transport systems integrated mangers, thus addressing and providing real time data of transport system, i.e., the busses, parking, traffic flows, etc. The system allows managing a big volume of data coming from a variety of sources considering both static and dynamic data, this data is then mapped to a Smart City Ontology and stored into an RDF-Store where this data are available for applications via SPARQL queries to provide new services to the users. The derived ontology has been obtained by means of an incremental process performed analyzing, integrating and validating each added data set. Thus the resulting ontology is a strong generalization of a large set of data modeling problems. In addition, a process of verification and validation have been deeply performed allowing to identify the set of triples to improve and enrich the model and the correction to be performed in order to enable the exploitation of the deductive capabilities of the final model. Finally, the proposed system also provides a visualization and exploration tool to explore the data available in the RDF-Store.

The next step will be to identify famous names, points of interest, locality names that can be linked to other data set as DBpedia[3] or GeoNames[4] according to a Linked Open Data model. This process can be performed with a simple NLP algorithm [6]. Furthermore an upcoming integration of the DISIT Ontology for Smart City with the GoodRelations model, is also planned, together with the automation of the reconciliation step, thanks to link discovery and machine learning techniques.

ACKNOWLEDGMENT

A sincere thanks to the public administrations that provided the huge data collected and to the Ministry to provide the funding for Sii-Mobility Smart City Project, a warm thanks to Lapo Bicchielli, Giovanni Ortolani, Francesco Tuveri.

REFERENCES

[1] Caragliu, A., Del Bo, C., Nijkamp, P. (2009), Smart cities in Europe, 3rd Central European Conference in Regional Science – CERS, Kosice (sk), 7-9 ottobre 2009.

[2] Bellini P., Di Claudio M., Nesi P., Rauch N., "Tassonomy and Review of Big Data Solutions Navigation", Big Data Computing To Be Published 26th July 2013 by Chapman and Hall/CRC

[3] Vilajosana, I. ; Llosa, J. ; Martinez, B. ; Domingo-Prieto, M. ; Angles, A., "Bootstrapping smart cities through a self-sustainable model based on big data flows", Communications Magazine, IEEE, Vol.51, n.6, 2013

[4] Ontology of Trasportation Networks, Deliverable A1-D4, Project REWERSE, 2005 http://rewerse.net/deliverables/m18/a1-d4.pdf

[5] Pan, Feng, and Jerry R. Hobbs. "Temporal Aggregates in OWL-Time." In FLAIRS Conference, vol. 5, pp. 560-565. 2005.

[6] Embley, David W., Douglas M. Campbell, Yuan S. Jiang, Stephen W. Liddle, Deryle W. Lonsdale, Y-K. Ng, and Randy D. Smith. "Conceptual-model-based data extraction from multiple-record Web pages." Data & Knowledge Engineering 31, no. 3 (1999): 227-251.

[7] Auer, Sören, Jens Lehmann, and Sebastian Hellmann. "Linkedgeodata: Adding a spatial dimension to the web of data." In The Semantic Web-ISWC 2009, pp. 731-746. Springer Berlin Heidelberg, 2009.

[8] Andrea Bellandi, Pierfrancesco Bellini, Antonio Cappuccio, Paolo Nesi, Gianni Pantaleo, Nadia Rauch, ASSISTED KNOWLEDGE BASE GENERATION, MANAGEMENT AND COMPETENCE RETRIEVAL, International Journal of Software Engineering and Knowledge Engineering, Vol.22, n.8, 2012

[9] Apache HBase: A Distributed Database for Large Datasets. The Apache Software Foundation, Los Angeles, CA. URL http://hbase.apache.org.

[10] Pentaho Data Integration, http://www.pentaho.com/product/data-integration

[11] Barry Bishop, Atanas Kiryakov, Damyan Ognyanoff, Ivan Peikov, Zdravko Tashev, Ruslan Velkov, "OWLIM: A family of scalable semantic repositories", Semantic Web Journal, Volume 2, Number 1 / 2011.

[12] S.Gupta, P.Szekely, C.Knoblock, A.Goel, M.Taheriyan, M.Muslea, "Karma: A System for Mapping Structured Sources into the Semantic Web", 9th Extended Semantic Web Conference (ESWC2012).

---

[3] http://dbpedia.org/
[4] http://www.geonames.org/